# Hybrid Finite Element and Material Point Method to Simulate Granular Column Collapse from Failure Initiation to Runout

Brent Sordo[1], Ellen Rathje, Ph.D.[2], Krishna Kumar, Ph.D.[3]

[1]Ph.D. Candidate, Department of Civil, Architectural, and Environmental Engineering, The University of Texas at Austin, 301 E Dean Keeton St, Austin, TX 78712; E-mail: bsordo@utexas.edu

[2]Professor, Department of Civil, Architectural, and Environmental Engineering, The University of Texas at Austin, 301 E Dean Keeton St, Austin, TX 78712; E-mail: e.rathje@mail.utexas.edu

[3]Assistant Professor, Department of Civil, Architectural, and Environmental Engineering, The University of Texas at Austin, 301 E Dean Keeton St, Austin, TX 78712; E-mail: krishnak@utexas.edu

## ABSTRACT

The performance evaluation of a potentially unstable slope involves two key components: the initiation of the slope failure and the post-failure runout. The Finite Element Method (FEM) excels at modeling the initiation of instability but quickly loses accuracy in modeling large-deformation problems due to mesh distortion. Hence, the FEM is unable to accurately model post-failure slope runout. Hybrid Eulerian-Lagrangian methods, such as the Material Point Method (MPM), offer a promising alternative for solving large-deformation problems, because particles can move freely across a background mesh, allowing for large deformation without computational issues. However, the use of moving material points in MPM for integration rather than the fixed Gauss points of the FEM reduces the accuracy of MPM in predicting stress distribution and thus failure initiation. We have created a hybrid method by initiating a failure simulation in FEM and subsequently transferring the coordinates, velocities, and stresses to MPM particles to model the runout behavior, combining the strength of both methods. We demonstrate the capability of the hybrid approach by simulating the collapse of a frictional granular column, comparing it to an empirical solution, and evaluating a suitable time to transfer from FEM to MPM by trialing multiple iterations with transfers at different stages of the collapse.

## INTRODUCTION

Landslides are potentially disastrous events that engineers take extensive precautions to anticipate and prevent. Predicting these natural hazards involves two key variables: predicting when/if an event will occur and understanding the magnitude and consequences of the event (i.e. the runout distance). Slopes are modeled through a variety of numerical methods to answer these questions, but most are focused towards the initiation of failure. One common method for modeling slope failures is the Finite Element Method (FEM), which is used to examine slope performance under various conditions and is very competent at evaluating when/if a failure will occur. However, the FEM is limited to solving small-strain problems (like the initiation of landslides) but quickly loses accuracy in large-deformation problems due to mesh distortion (Cuomo et al., 2021). This issue critically restricts the utility of the FEM for modeling landslide runout that can extend to hundreds of meters.

The Material Point Method (MPM) (Bardenhagen et al., 2000) offers a solution to this issue. A hybrid Eulerian-Lagrangian approach, MPM models the soil mass as individual material points that can move and deform according to Newtonian laws without distorting the background mesh. Thus, MPM can account for large displacements without losing accuracy. This characteristic enables the accurate evaluation of the landslide runout (Cuomo et al., 2021). Runout extent is an important consideration for evaluating the consequences of slope failure, and MPM can effectively predict it. However, integration of stresses on the moving material points results in loss of accuracy in MPM. MPM stress states are less precise and prone to checkerboarding, and absorbing boundary conditions (and thus seismic problems) are difficult to model due to particle movement away from the model boundary. Therefore, MPM is less competent at modeling the initial state and complex triggering mechanisms, such as earthquake-induced liquefaction. Thus, we propose a hybrid FEM-MPM approach to capture the entire runout process.

In this paper, we present a novel FEM-MPM hybrid method that simulates landslide initiation via FEM and then landslide runout via MPM, combining the advantages of both numerical methods and evaluating the slope performance with a single model. To our knowledge, FEM and MPM have been combined previously to operate side by side governing different sections of a model (Lian et al., 2011; Pan et al., 2021) and separately to model the same failure and then directly compared (Cuomo et al., 2021), but not in sequence over the same model. We utilize the open-source FEM program OpenSees (McKenna, 1997) for modeling the failure initiation, and the CB-Geo MPM code (Kumar et al., 2019) for the runout. To demonstrate the validity of this FEM-MPM hybrid method, we simulate the granular column collapse experiment (Lajeunesse et al., 2005; Soundararajan, 2015), in which the runout distance is a function of its initial geometry (aspect ratio of the column). This column collapse is modeled by FEM, MPM, and our FEM-MPM hybrid, so each models' results can be directly compared to the known solution. There are also multiple iterations of the FEM-MPM hybrid model to determine the ideal time to transfer between the two methods.

**COLUMN MODEL**

This granular column collapse experiment involves creating a 50 m wide and 40 m tall column of dry, frictional soil with an aspect ratio (H/L) of 0.8 with a frictional surface at its base and a frictionless roller boundary on its left. Initially, the column is assigned a linear elastic constitutive model with a Young's modulus of 172 MPa, Poisson's ratio of 0.23, and mass density of 1925 kg/m$^3$, establishing an initial, *in situ* equilibrium stress state under geostatic conditions. Then, the constitutive model is changed to a Mohr-Coulomb material with $\phi= 22.2°$ and minimal cohesion, matching the soil modeled by Kumar and Soga (2019). This change in the constitutive model simulates the opening of a gate on the right side, allowing the soil to fail and resulting in a runout. Figure 1 displays the initial geometry, elastic stress state, and fixities of the model in OpenSees.

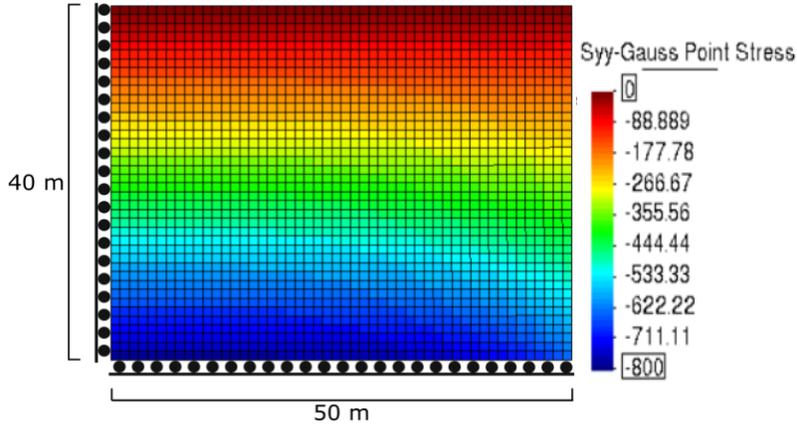

**Figure 1: Initial state of model in FEM (stress in kPa and negative indicates compression)**

We simulate the column collapse via three different configurations: (a) FEM only, (b) MPM only, and (c) hybrid FEM-MPM at different transfer times. The FEM model employs linear, square 1m x 1m finite elements with a single Gauss point and four nodes (McGann et al., 2012), and the MPM model utilizes the Generalized Interpolation Material Point Method (Bardenhagen et al., 2004), also with a 1 m x 1 m resolution mesh, and sixteen material points per mesh cell. The runout predictions from the pure FEM and pure MPM models serve as a baseline to which we compare the hybrid approach. In the hybrid FEM-MPM method, we extract the FEM stress, position, and velocity data at the nodes at a given time and utilize them as the inputs to the material points in an MPM analysis for runout. We use a Python script to convert the ASCII data from OpenSees to HDF5 files, which can be resumed in the MPM code.

**HYBRID FEM-MPM**

The transfer from FEM to MPM is shown works according to the workflow in Figure 2. During the transfer, we divide each FEM element into sixteen material points for the MPM analysis, placed at the Gauss point locations of each element for 4 x 4 numerical integration (Figure 3). We approximate the stress at each node by averaging the values of stress at the Gauss points of each adjacent element, weighted by the inverse of the distance between the node and each Gauss point. Then, we interpolate the initial MPM particle stresses, coordinates, and velocities based on the respective values of the nodes via the elements' shape functions. Finally, we calculate FEM element volumes based on their respective distorted or current element dimensions and distribute that volume among the material points according to the associated Gauss weights. The mass of each particle is assigned based on its volume and mass density.

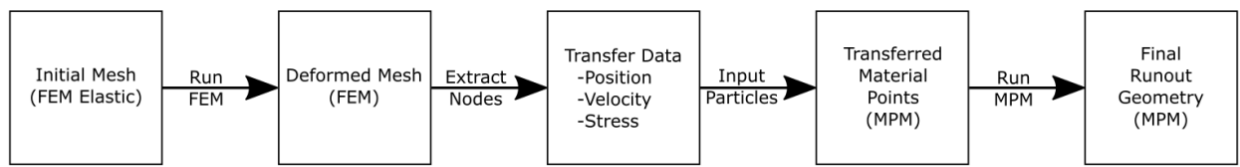

**Figure 2: FEM to MPM Transfer Workflow**

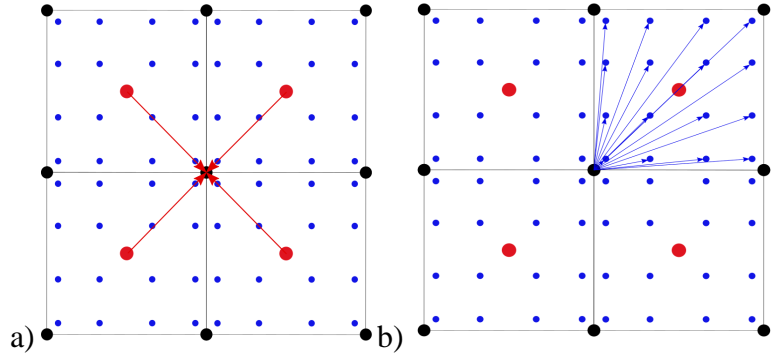

**Figure 3: Example FEM-MPM transfer geometry. (a) Transfer of stresses from Gauss points (red) to nodes (black) and (b) from nodes to particles (blue)**

We facilitated transfers from FEM to MPM at $t = 0$ s (immediately after the elastic phase of the FEM model) as well as at $t = 0.5, 1, 2,$ and $3$ s of FEM model time after the change in constitutive model. Transfers at later times proved to be unusable as the FEM mesh was too severely distorted. Figure 4 displays the four stages of this FEM-MPM hybrid model for the $t = 2$ s transfer, with the FEM mesh after the elastic phase and at the $t = 2$ s transfer time and the MPM material points immediately after the transfer and at the new equilibrium state when $t = 20$ s.

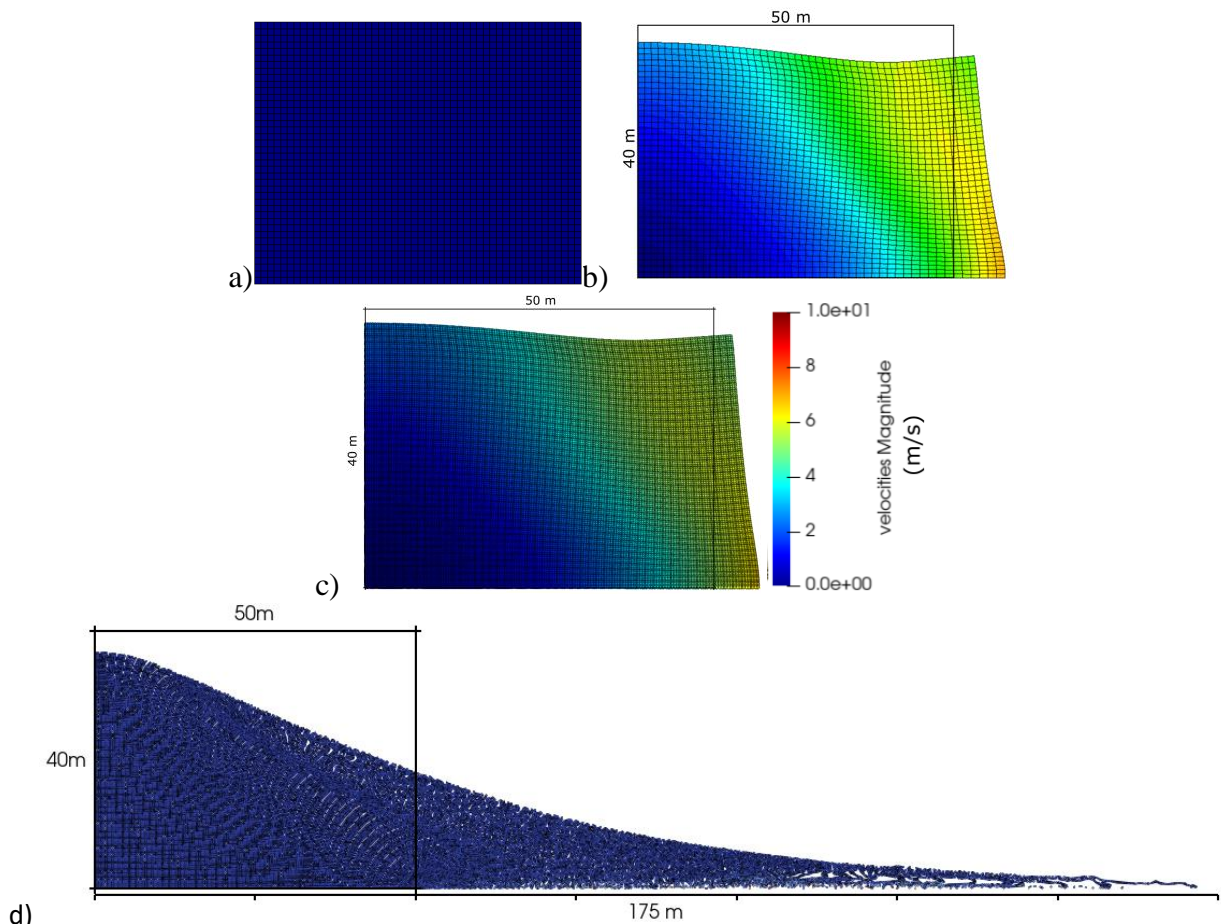

**Figure 4: Velocity (m/s) contours of stages of FEM-MPM transfer. (a) FEM elastic model. (b) $t = 2$ s in FEM model. (c) $t = 2$ s in MPM. (d) $t = 20$ s in MPM model**

## RUNOUT RESULTS

We evaluate the results of the column collapse similarly to Kumar and Soga (2019). We define runout as the difference in the length of the model (i.e., $L - L_0$ where $L$ and $L_0$ are the current and initial column lengths, respectively, normalized by initial length ($L_0$):

$$normalized\ runout = \frac{L-L_0}{L_0}.$$

The normalized height is also considered to evaluate the extent to which the highest point of the column has fallen. Normalized height is defined as the ratio of the current height ($H$) and initial height ($H_0$):

$$normalized\ height = \frac{H}{H_0}.$$

$L$ is defined as the 98$^{th}$ percentile particle/node to the right, and $H$ as the 99.5$^{th}$ percentile particle/node to the top to avoid spurious individual particles which might separate from the model. $L_0$ and $H_0$ are 50 m and 40 m, respectively.

We considered the evolution of these two parameters as a function of time normalized by the critical time ($\tau_c$; Lajeunesse, 2005), the time when the flow is fully mobilized, defined as:

$$normalized\ time\ = \frac{t}{\tau_c} = \frac{t}{\sqrt{\frac{H_0}{g}}}.$$

For $H_0$ of 40 m, $\tau_c$ is 2.02 s.

Figure 5 displays the evolution of the surface profile of the purely FEM and the $t = 1$ s transfer FEM-MPM hybrid model, featuring contours from $t = 1$, 3, and 5 s (i.e., t / $\tau_c$ ~ 0.5, 1.5, and 2.5). We run the FEM model for only 5 s (t/ $\tau_c$ = 2.5), beyond which FEM experiences excessive mesh distortions and the simulations were unusable, which is described in the FEM Deformation Response section. The surface profiles in Figure 5 reveal that the FEM column height decreases much faster than the FEM-MPM hybrid model, but the runout of the FEM-MPM hybrid model extends faster and further. Furthermore, the corner of the FEM model persists rather than smoothing, and the elements near the toe begin folding over each other and fall below ground level by $t = 5.0$ s. The surface geometry predicted by the FEM-MPM hybrid model approaches a smoother, more realistic final geometry that is similar to soil settling to an angle of repose.

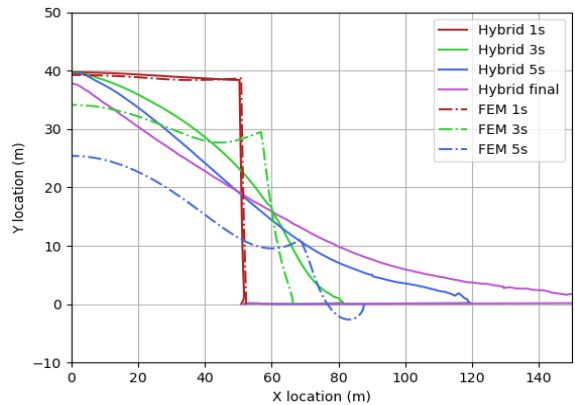

**Figure 5: Surface evolutions of Hybrid (1s Transfer) and FEM between 1s and 5s**

Figure 6 displays the time evolution of the normalized runout and normalized height of column collapse for the pure MPM, pure FEM, and hybrid FEM-MPM models transferred at different times. Consistent with Figure 5, Figure 6b shows that the height falls faster in the FEM model and Figure 6a demonstrates that the runout moves faster in the FEM-MPM hybrid and purely MPM models. This trend is the same for the pure MPM model and all hybrid FEM-MPM models independent of the transfer time. Furthermore, the FEM model never settles into a new equilibrium, while the MPM and FEM-MPM hybrid models eventually do reach a new state of static equilibrium. The FEM results are only plotted up to a normalized time $t / \tau_c = 1.75$, beyond which pure FEM results are too unrealistic due to mesh distortions. The pure MPM and hybrid FEM-MPM models behave similarly, but when the transfer occurs at $t = 2.0$ s ($t / \tau_c \sim 1$) or later, the final geometry begins to differ, especially the height. Except for the $t = 3.0$ s ($t / \tau_c \sim 1.5$) transfer, the final normalized runout of about 1.6 is very close to the empirical model of Soundararajan (2015), which was 1.53 for the aspect ratio of our column.

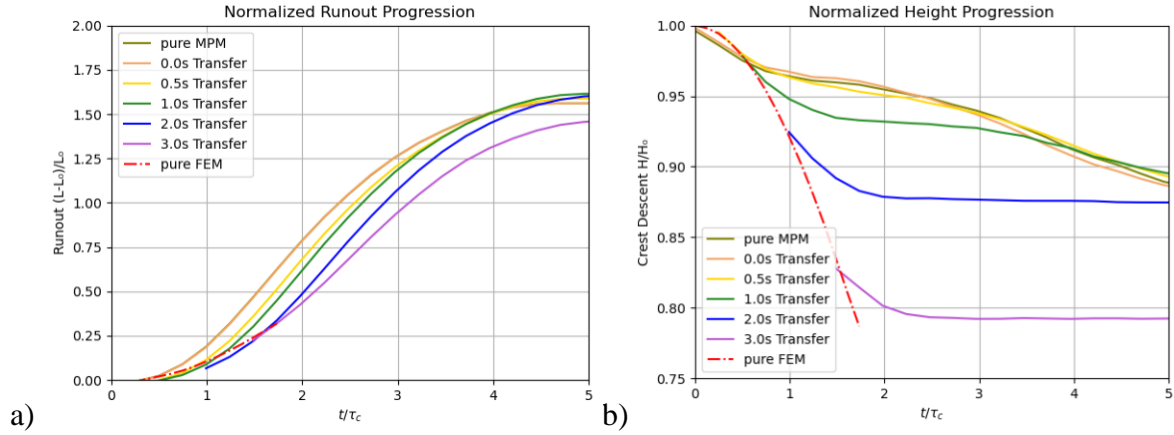

**Figure 6: (a) Normalized Runout and (b) crest descent of all models**

Figure 7 displays the evolution of potential energy (PE) and kinetic energies (x and y-components) normalized by the initial potential energy. We determined these quantities by calculating the aggregate energy of each particle (for MPM) or element (for FEM):

$$PE = \sum_{i=1}^{n_{particles}} mass \times g \times height$$

$$KE_x = \sum_{i=1}^{n_{particles}} \frac{1}{2} * mass \times velocity_x$$

$$KE_y = \sum_{i=1}^{n_{particles}} \frac{1}{2} \times mass \times velocity_y$$

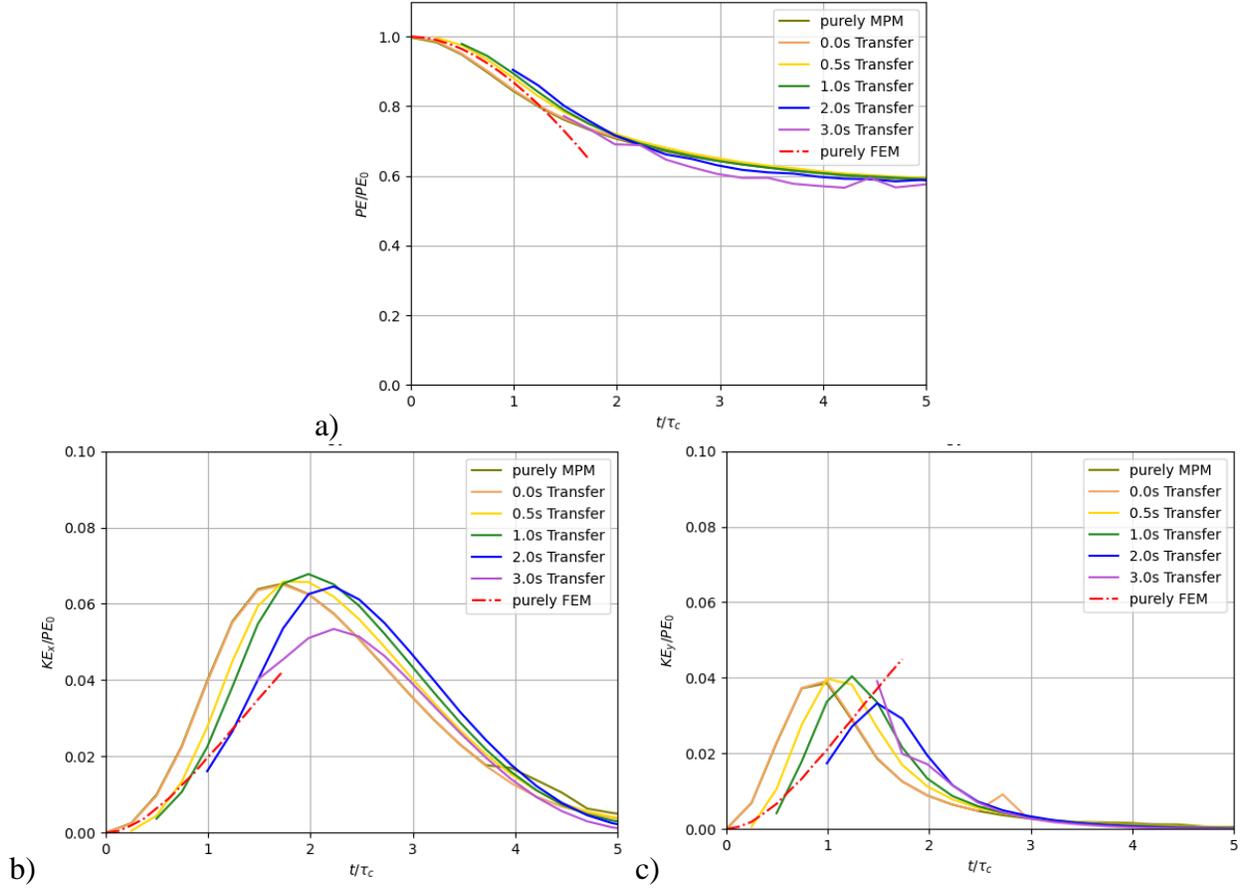

**Figure 7: (a) Evolutions of total potential energy, (b) X-component kinetic energy, and (c) Y-component kinetic energy across all models, normalized by initial total potential energy**

Potential energy decreases during the column collapse (Figure 7a) and eventually stabilizes to a new constant value of ~ 0.6 the initial value. The kinetic energy reaches a peak during the collapse and then resettles to zero after movement has ceased (Figures 7b and 7c), although the peak of the y component occurs significantly sooner than the peak of the x component, indicating that the runout is dominated first by vertical motion and then by horizontal motion. As in Figure 6, the FEM-MPM hybrid models and the pure-MPM model behave similarly but the formers begin to differ, with peaks occurring later, by approximately half the duration of time between the transfers. For the FEM results, neither component of kinetic energy shows any sign of stabilizing. In fact, the continued increase in kinetic energy indicates continuous acceleration even after the pure-MPM and hybrid FEM-MPM models have begun to decelerate. The stabilization of the kinetic energy, as well as of the runout and height, serves as an indicator that the column collapse has ended and provides insight into the manner in which the collapse occurs. Once again, this result demonstrates that runout stabilizes to a new equilibrium in the MPM analysis but does not in the FEM analysis.

Soundararajan (2015) reported that when $t/\tau_c = 1$, the column collapse had fully developed and by $t/\tau_c = 3$, the flows came to rest, but the movement in the models from this study appears to be slower than the results reported by Soundararajan (2015). For the pure MPM model, the vertical component of kinetic energy reaches its peak at $t/\tau_c \sim 1$, but the horizontal component reaches its peak at $t / \tau_c \sim 1.75$. The hybrid FEM-MPM models reach similar peaks but up to 0.5 units of normalized time later, depending on the time of the transfer. The collapse does not come to rest until after $t / \tau_c = 5$ regardless of transfer time.

**FEM DEFORMATION RESPONSE**

Figure 8a shows the deformed FEM geometry along with contours of shear strain from the FEM at $t = 5.0$ s ($t / \tau_c = 2.5$). By this time, the maximum shear strains exceed 100%, resulting in severely deformed elements. Note that the elements have begun to fold over themselves and even dip below the ground level in the bottom right corner, clearly displaying the unrealistic results that are delivered when the mesh is so distorted. To quantify the deformation of the mesh, we calculated the minimum Jacobian determinant across all elements as a function of time (Figure 8b). For $t / \tau_c > 1.75$, the minimum has fallen below zero, contributing to serious numerical issues, but degradation of the Jacobian begins by $t / \tau_c > 0.5$, so a transfer should ideally be performed before that time.

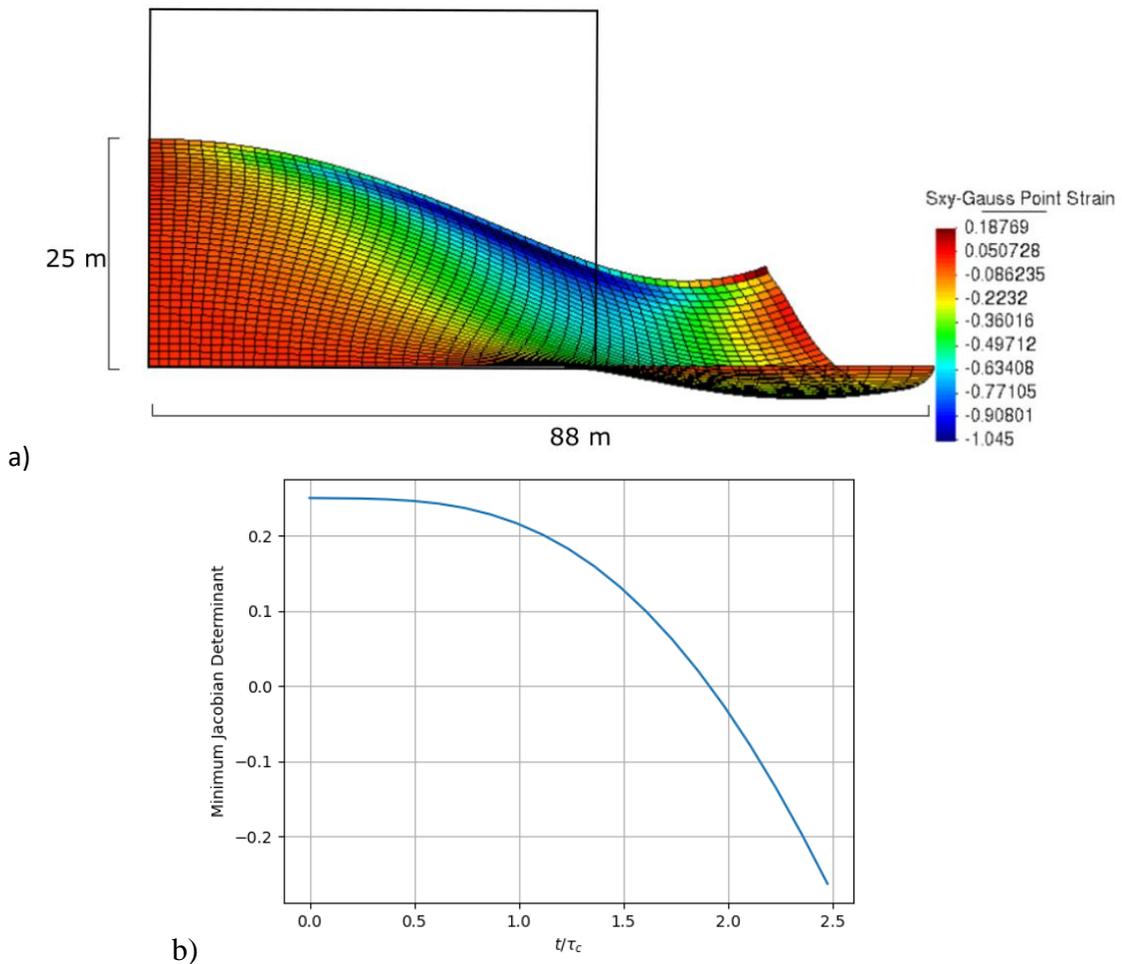

a)

b)

**Figure 8: (a) Deformed FEM mesh at $t = 5$ s and (b) evolution of determinant of minimum Jacobian with time**

**MPM STRESS RESPONSE**

The MPM models all suffer from severe stress checkerboarding, across all types of stress, timesteps, and transfer times. Figure 9 displays an example of this phenomenon, in which individual particles within the model experience anomalously large values of shear stress in a repeating pattern. With the contour intervals set to the range of the FEM stresses, the many particles which are under extreme stresses become apparent along with their checkerboard pattern, and these extremes can be orders of magnitude greater than the FEM stresses. This effect is immediate after the transfer from FEM to MPM, regardless of when the transfer is initiated. This stress checkerboard pattern is not realistic and is a deficiency of the MPM modeling procedure, but the effects do not seem to be critical to the behavior of the runout in the column collapse. However, other models which include pore pressures and liquefaction would be more sensitive to its effects. This problem does not occur in the FEM models, which produce much smoother stress distributions.

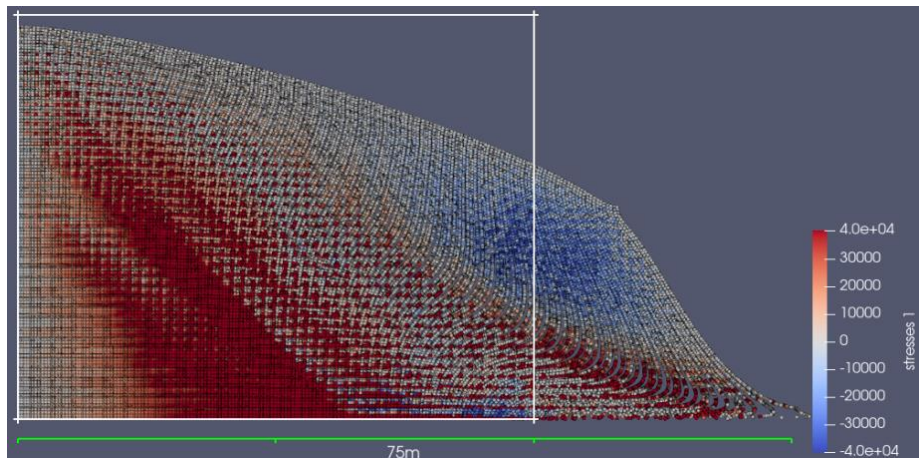

**Figure 9: Checkerboarded shear stress distribution of FEM-MPM hybrid model at $t$ = 3 s ($t$ / $\tau_c$ ~ 1.5) after transfer at $t$ = 1 s ($t$ / $\tau_c$ ~ 0.5)**

**SUMMARY**

We develop a hybrid FEM-MPM approach to model slope failures from initiation to runout response. We evaluate the hybrid approach by simulating a granular column collapse experiment and comparing the results with pure-FEM and pure-MPM simulations. The pure-FEM is initially very precise but loses accuracy quickly and delivers unrealistic results after the mesh begins to deform. On the other hand, the pure-MPM delivers more realistic runout behavior but suffers from checkerboarding of stresses. In the hybrid FEM-MPM approach, we transfer the coordinates, velocities, and stresses from FEM to MPM at various times to model the entire runout process. The results match the empirical solution documented by Soundararajan (2015) relatively well, establishing the viability of this FEM-MPM hybrid method.

We consider a few iterations of this FEM-MPM hybrid model by facilitating the transfer at five different times to identify an ideal. The quality of the transfer model appears to suffer if the transfer is facilitated too late, as it will carry with it some of the unrealistic deformation of the FEM. Therefore, the transfer should ideally occur before the minimum Jacobian determinant significantly decreases. From this experiment, there do not appear to be any potential issues from transferring too soon, though MPM's the stress checkerboarding may cause issues in other scenarios.

The runout behavior of granular column collapse is simply a function of its initial aspect ratio. Hence, the inaccuracies of stresses in the initial states of the MPM do not lead to significant differences in the runout behavior. However, in problems where the failure initiation is governed by the initial stresses, such as earthquake-induced liquefaction, it is important to accurately capture the stress response. If pore pressure is a consideration, exact stress distributions will be a determining factor for where and when liquefaction will occur, a phenomenon which fundamentally alters the behavior of the soil. In this case, we would model the problem in FEM long enough to accurately capture the onset of liquefaction and then transfer to MPM to simulate the runout response.

This MPM-FEM hybrid regime promises applications that neither FEM nor MPM can execute independently. Properly utilized, it may be able to fully model the onset and runout of complex landslides resulting from seismically induced liquefaction, among many other applications which formerly could not be fully modeled by either FEM or MPM.